\documentstyle{article}
\begin{document} 
\title{Anisotropy of CMB \& Cosmological Model}
\author{V.N. Lukash\\
	Astro Space Center of Lebedev Physical Institute\\
	Profsoyuznaya 84/32, 117810 Moscow Russia}

\maketitle

\abstract{
In this brief review I will touch upon the main points 
and trends of observational and theoretical cosmology, that 
influence and change today our understanding of the Universe.
}

\section{Introduction}

My goal is to highlight current situation in building up the
true cosmological model. Nowadays, after 30 years of the 
discovery of the CMB relic radiation, its anisotropy is revealed 
and provides a basic channel of our information about the World. 
If to be brief, the cosmology is determined today up to the 
accuracy 20-30 \% which is a great progress since what we had 
10-20 years ago when the discussions at best were on the level 
of a factor of two or even more. The great hope of cosmologists 
is related with the development of ground based and space based 
$\Delta T/T$ experiments which will be able to delimit and fix
the cosmological model up to few per cent in the nearest future.

\section{Basics of the Problem of LSS Formation}

The seeds of the visible {\it Large Scale Structure} in the
Universe are the {\it Cosmological Density Perturbations} which
grow due to the gravitational instability in the late, cold
period of the Universe expansion when the density is dominated by
{\it Cold Dark Matter} ($z < 10^5$). These primordial CDPs must
have been created since the inflationary Big Bang epoch, as after
the end of inflation -- the inflaton decay and the 
primeval reheating of cosmic medium -- the Universe became
predominant by relativistic particles and radiation which 
remnant today as the fossil CMB. Such an early hot Universe 
was absolutely gravitationally stable
against small perturbations of the matter density and
gravitation field: the CDPs were a kind of frozen, just
presenting the 'gravitating sound waves' propagating across the
relativistic matter with a constant amplitude. Thus, the
CDPs started growing only after the equality epoch when the
pressure diminished in comparison with the third fraction of the matter
density. The required CDP amplitude for galaxy clusters could
form by now, has therefore been predicted to consist $\delta \sim
10^{-5}$, which was finally confirmed by COBE\cite{COBE} on
this same level but the two orders of magnitude larger scale
(that in turn appeared to be in a successful consistency with
another famous prediction known as the {\it Harrison-Zel'dovich}
scale-invariant perturbation spectrum).

This optimistic situation has created a great impetus for the
observational and theoretical cosmology extending dramatically
by continuing progress in the improved technology of deep-sky surveys and
CMB temperature detections. The ultimate goal has come for the
reconstruction of the primordial CDP spectrum from 1 Mpc
scales up to the horizon, the scope straighforwardly related to the high
energy physics at inflation capable by observational
tests today.

Three points should be emphasised in connection with the problem
of LSS formation: {\it Theoretical, Model, and Observational}. The
first point means that the creation of the LSS in the Universe is 
as fundamental problem as the creation of the Universe as whole: 
both features, the small CDPs and the background Friedmann model 
(the {\it Cosmological Principle}), are produced in the process of 
inflation in the very early Universe. The theory works at very high 
energies ($\sim 10^{16}$ GeV) whereas the observations proceed in 
the low-energy limit ($\sim 10^{-4}$ eV). To provide a fair 
comparison in such a situation one needs a model to know how the 
perturbations evolved during the whole history of the Universe.
Any confrontation in cosmology between theory and observations is 
thus model dependent.

To fix the model one needs priorities and the fundamental
constants.  The former conventionally assume the gravitational
instability mechanism as a principal tool for the CDP dynamics
on large scale and the Gaussian primordial perturbations (with
random spatial phase).  The latter requires the knowledge of the
current time when the structure is observed ($H_0$), the
abundance of cosmic matter components ($\Omega_M$,
$\Omega_\Lambda$, $\Omega_b$) and {\it Cosmic Gravitational
Waves} (T/S), and the nature of the dark matter (e.g. relic
scale field, CDM, HDM, the number of species of massive
neutrinos and relativistic particles, etc.).  Today,
cosmologists venture the following approach: if the dark matter
model is stated as fairly simple (with small number of the
parameters, e.g. the one with zero spatial curvature or zero
$\Lambda$-term, or other) then the recovering of both the CDP
spectrum and the model fundamental parameters, can be provided
by observational data on $\Delta T/T$ and LSS.

Below, I come to discussion of the models in such a conventional
probability sense. However, there is no principal restrictions:
any theory could be tested to the limit if we have enough data.
The more data will be available, the less uncertainties will
remain in the model and more parameters can be finally
determined.  Actually, we are in the beginning on the way of
getting the data. Today we may do a model restoration exercise
facing the observations we just have. 

\section{Dark Matter Models}

Until recently there were two basic theories claiming to
approach the corner stones of the LSS formation: inflation and
defects. While being very much different in their grounds on
galaxy seeds -- the linear Gaussian scalar perturbations in the
one case and the non-linear non-Gaussian cosmic defects
(strings, monopoles, textures) in the second scenario -- both
models presented the fundamental inevitable perturbations
produced by the physics of the very early Universe: the
parametrically amplified quantum vacuum fluctuations of the
inflaton and the topological defects left after phase
transitions, respectively.

However, the defect model normalised by the CMB fluctuations
proved to fail to meet the LSS formation\cite{defect}. The
reason is that the non-linear matter perturbations generate all
three types of the metric fluctuations - {\it Scalar, Vortex}
and {\it Tensor} ones, which all contribute to the Sachs-Wolfe
$\Delta T/T$ anisotropy on large angular scale, so the residual
S-mode amplitude appeared to have no sufficient power to develop
the observed galaxy distribution.

At the moment, only the inflation theories got through ordeals
fitting the LSS and $\Delta T/T$ observations. The principal
quest here is the predicted Gaussian nature of small CDPs, which
faces a satisfactory consistency with the real distribution of
galaxies on scales $\sim 20\; h^{-1}$ Mpc\cite{dist}. The only
obstacle to test reliably this important feature of the CDP
seeds is the limited depth of the available galaxy surveys.

The deep galaxy surveys would be highly welcome to solve also
other challenge of the modern cosmology: the fractal model
attacking persistently the cosmological principle. The point is
that the huge voids seen in the galaxy distributions extend up
to scales $\sim 100\; h^{- 1}$ Mpc which is close to the
catalogues' sizes, thus, leaving a room for discussions on the
homogeneity scale\cite{void}. Nevertheless, I would like to
stress that the fractal challenge is still a question for
distribution of the baryonic matter (the optical galaxies)
rather than for the total mass of the Universe. The latter
should be pretty homogeneous on scale larger than few tens of
Mpc to fit the beautiful Hubble diagrams. Not to speak on the
uniform microwave and X-ray backgrounds evidencing the cosmic
homogeneity on larger scales.

Thus, we consider only models backed on the inflationary
theories. The main tool for the Gaussian perturbations is the
second moment of their distribution related to the power-spectrum:

\begin{equation}
\langle \delta^2 \rangle =
\int\limits_0^{\infty}P(k) k^3 dk =
\int\limits_0^{\infty} \Delta_k^2\frac{dk}{k}.
\end{equation}

The dimensionless CDP spectrum $\Delta_k^2$ has a simple meaning
of the variance of density contrast in the scale $k$ (the wave
number) within the scale band $dk \sim k$, it is evidently
additive ($\delta^2 \sim \Sigma \Delta_k^2$).

Before coming to
discussion on the spectrum observational reconstruction let me
sketch briefly the situation with the model parameters.

\section{Cosmological Parameters}

It seems that the long strong debate on $H_0$ is approaching
to its end and we are going to learn the value of the Hubble
constant during nearest years. Today, very premising seem two
methods: measuring Cepheids in distant galaxies and the
supernovae type Ia method. I would not like to fix here the
number since it is not yet time for any consensus between the
groups about systematic and selection bias effects for all
methods employed.  For us, it is important to note that standard
cosmological models (with the critical dynamical matter density,
$\Omega_M = 1$, and negligible $\Lambda$-term) are consistent
only with small Hubble constant $(H_0 < 65\, km\,s^{-1}
Mpc^{-1})$ regarding the low limit for the age of the Universe
coming from globular clusters.

A much more optimistic point stands on the current progress in
determination of $\Omega_M$ and the baryonic content of the Universe
$\Omega_{b}$. At the first glance the situation looks similar:
again we have two groups of experiment resulting in different
conclusions. However, here the consensus is possible.

The first experiment deals with megaparsec scales -- galaxy
halos, groups and X-ray clusters, -- $l<l_{D}$ (the dynamical
scale in the Universe $l_{D}\sim 10 h^{-1} Mpc$, which is the
largest pass of a galaxy for the current Hubble time, just the
scale of the richest collapsing clusters). 
The assumption on the hydrostatic equilibrium yields
a low dynamical mass responsible for the formation of the
gravitational potential on small scales: $\Omega_M \sim 0.3$.
Another important observation is large fraction of baryons inside
X-ray clusters reaching somehow $\sim 20 \%$ within scale $\sim 1$ Mpc:
\begin{equation}
\frac{M_{b}} {M_M} \sim 0.2, 
\end{equation}
which is also consistent with a low matter density involved dynamically
in megaparsec scales (as $\Omega_{b} \le 0.1$ due to the
primordial nucleasynthesis and $M_{b}/M_M$ may be $\sim
\Omega_{b}/\Omega_M$ on the dynamical scale).

The other experiment deals with LSS ($\ell > \ell_D$) and argues 
that the 
Universe may be matter dominated, $\Omega_M \simeq 1$. There are few 
principal arguments for this
(still a more detail model dependent ones in comparison with the
small scale arguments):
\begin{itemize}
\item
the existence of  substructures in the majority of galaxy clusters
evidencing that the clusters are just forming systems which is
possible only in the Universe dynamical close to the critical density, 
\item
the large coherence velocities obviously of the cosmological
origin, thus allowing for the reconstruction of the total density
contrast (and as a consequence, the consistency with the "standard" 
model $\Omega_M \simeq 1$ and the galaxy biasing factor  $b \simeq 1$),
\item
the fact that the Gaussian nature of the linear primordial
cosmological perturbations may be recovered (after going back in 
time from the actual non-linear distribution of matter density 
and velocity) only for the parabolically expanding Universe 
$(\Omega_M \sim 1)$,
\item
the weak gravitational lensing confirming high dynamical mass 
abundance around some X-ray clusters,
\item
the lensing argument on the fraction of splitting quasars, 
still much dependent on the model assumptions,
\item
the evolutionary argument on the galaxy clusters abundance,
\item
the geometrical argument from the distant supernovae type Ia,
\item
the point coming from $\Delta T/T$ anisotropy (mainly, the 
location of the first acoustic peak).
\end{itemize}

The last three points got some dramatic turns in the recent
time which I cannot help mentioning here.

It is the ENACS identification of the nearby galaxy clusters by
the dispersion velocities of their optical galaxies\cite{bum}
that has shown the previous underestimate of the Abell
cluster abundance (which employed the method of counting the number
galaxy concentrations in the sky). At the moment we may state 
the consistency of the cluster number density evolution in
redshifts with the $\Omega_M = 1$ Universe. At least, the low
evolution argument that for many years was considered as a basic 
argument in favour of the low density Universe, is not any more 
as strong as it seemed before.

The breakthrough in the problem of the model geometry
restoration is being done today by the classical Hubble diagrams
(the redshifts {\it vs} apparent magnitudes) composed from
distant supernovae of type Ia\cite{SN}. Contrary to galaxies,
such sources look amazingly standard candles which is well
supported by the distance measurements to nearby supernovae. The
candidate model for this class of supernovae could be exploding
white dwarfs that would clearly fix their mass and chemical
composition and thus provide theoretical understanding of their
observable luminosity stability. Tested by the distant
supernovae, the deviations of the Hubble diagram from the linear
law hint on the real geometry of the Universe. While more
statistics is still needed to come to the reliable conclusions,
the first {\it conditional} estimates of the model parameters
seem very much impressive: if $\Lambda = 0$ then $\Omega_M <
0.6$, whereas for the flat models $\Omega_\Lambda < 0.5$.  The
latter inequality is especially interesting as it improves
dramatically the best lensing upper limit for the $\Lambda$-term
($\Omega_\Lambda < 0.6$).

The reconstruction of the cosmological parameters from CMB
temperature fluctuations recall today an exercise as the
strongest effect comes from the location and amplitude of the
'Doppler peak' (the first Sakharov oscillation) whose
observational detection leaves something to be desired. However,
not discussing here the numbers, it is worth while remembering
the {\it tendency} for the model parameters limits constrained
by all the $\Delta T/T$ data available in the
literature\cite{ssil}: it is for low $H_0$ ($\sim 0.5$) and high
$\Omega_b$ ($\sim 0.1$) and $\Omega_M$ (consistent with the flat
Universe) that comes somehow larger than the local astrophysical
predictions. The low density Universe ($\Omega_M < 0.3$) is
rejected by current $\Delta T/T$ data. 

Finally, a possible reconciliation between the DM experiments on
small and large scales can be the following: some fraction of dark
matter in the Universe is distributed on large scales and does
not enter the galaxy halos and groups. How it can be arranged?
Today we have purely theoretical ideas on such model.
The most frequently discussed are that with mixed dark matter
(hot+cold, with the hot particles like massive neutrinos with a
few eV rest mass and the corresponding density parameter
$\Omega_{\nu} \in (0.2, 0.4)$), and the model with non-zero
$\Lambda$-term $(\Omega_{\Lambda} \in (0.6, 0.7))$. For both
cases, the cold particles form the dynamical structure beginning
from small scales while on the large scale there is additional
contribution coming from light neutrinos or vacuum density (the
$\Lambda$-term), respectively. A very sceptical point concerning
these and other cosmological models considered currently as possible
candidates for the real Universe is as follows: all they are
multi-parameter and thus non-fundamental models, which differ
them drastically from the purely hot or purely cold dark matter
models.

Is there something very important which we miss in our
discussion on the formation of the Universe structure? May be. I
can only conclude here saying than none of the models under
discussion meets all the observational tests. Say, regarding the
two previous examples, for $\Lambda \ne 0$ models one can expect
a large fraction of old (relaxed) galaxy clusters and lensed
quasars whereas the hot+cold models require $H_0 < 65\,
km\,s^{-1} Mpc^{-1}$ and too a small abundance of X-ray clusters
and high-redshift quasars. Probably, the dark matter can exist
in the form of the relic scalar field left after inflation or in
some other exotic form, which require a more detail analysis.

In such a situation, the observational verifications become
extremely important. The principal tests is LSS of the early
Universe.

\section{The spectrum of density perturbations}

The cosmological models of LSS formation discussed today are
aimed to fit the observational data at $z = 0$ . So, we cannot
distinguish between them without going to the evolution at
medium and high redshifts where the models demonstrate their
essential difference. Two main experiments promote a snow ball
progress in the reconstruction of CDP spectrum, that was
impossible in previous years: $\Delta T/T (\theta > 1^\prime)$
and direct investigation of the evolution and hierarchy of LSSs.
The reason for stimulating such a progress is that these two
experiments confront and overlap each other: the $\Delta T/T$
investigations go nowadays to small comoving scales up to $\ell
\sim 10\;h^{-1}$ Mpc (recall the corresponding angular scale in 
arcmin $\theta \sim \ell h$), at the same time we observe a 
developed structure of clusters, filaments, voids, and 
superclusters coming up to scales $\sim 100\; h^{-1}$ Mpc. 

Any reasonable assumption on the "formation" of large voids and
superclusters in Gaussian perturbation theories inevitably leads
to $\Delta T/T$ predictions at $\sim 1^0$ capable of current
detection. There is a great puzzle that namely this scale
specifies the horizon at the decoupling era, therefore, the
angular scale of the first acoustic peaks. Its existence was
predicted by the theory long ago. Now, the time came for the
observations: it is just on agenda, a matter of the improved
instrument's technology and foreground separations that will
precisely determine the peak parameters and ultimately prove and
fix the theory.

Today, we are aware of the cosmological temperature anisotropy
on large scale and have some information on the whole spectrum
of the CMB fluctuations\cite{dtt}. Fortunately, the small
angular scales ($\theta < 1^0$) can be effectively tested from
the ground. The hope is that such the ground based instruments
as SK, CAT, VSI, RATAN-600, together with the balloon
experiments as well as the MAP, RELICT2 and Planck Surveyor
satellites will provide an advance sensitivity to put the point 
in the cosmology model reconstruction.

Meanwhile, the situation with the CDP spectrum  looks rather 
dramatic. On  large scale ($\sim 1000\; h^{-1}$ Mpc) the 
fundamental spectrum is small in amplitude and consistent with 
the HZ one:
\begin{equation}
\Delta_k^2 \sim k^{3 + n_S},\;\;\;\;n_S = 1.1\pm 0.1.
\end{equation}
However, in close scales ($\le 100 \; h^{-1}$ Mpc) the power
should be boosted as we observe a rich structure in the spatial
distribution of galaxies, clusters, $Ly_{\alpha}$-forest, and
distant sources like quasars. The latter is especially
important. We live in the period of the decay of quasar and
star formation activity\cite{9}. So, we have a unique
opportunity to use these numerous early sources to observe the
past dynamics of the LSS formation. This would be extremely
informative as the LSS perturbation amplitude, being still small
today at $\ell \sim 100\;h^{-1}$ Mpc, were even lower in the
past, which predicts a strong inverse evolution of such huge
systems as superclusters and voids.

It seems that quasars, the active galactic nuclei of distant
galaxies, form the LSS at medium redshifts ($z \sim 1-2$) which
is provided by their correlation function and the existence of
huge QSO groups recalling in properties (the comoving size and
abundance) local superclusters\cite{10}. Actually, distant 
bright quasars
may originate in merging galaxies in protoclusters, and thus can
trace the sites of enhanced matter density at medium and high
redshifts analogous to how galaxy clusters trace them in nearly
space. If so, then the dynamical formation of these early LSSs
suggests that the spectral amplitude on superclusters scale
($\sim 100\; h^{-1}$ Mpc) should be comparable and pretty close
to that on cluster scale ($\sim 10 \; h^{-1}$ Mpc), i.e. the CDP
spectrum is nearly flat between those scales\cite{11}:
\begin{equation}
\Delta_k^2 \sim k^{0.9 \pm 0.2}.
\end{equation}
This estimate for the spectrum slope is also indicated by the local 
observations of galaxy and galaxy cluster distributions\cite{12} 
$,$\cite{13} $,$\cite{14}.

A drastic break in the spectrum slope from the HZ asymptotic to
the flat part (4) should have happened at supercluster scale
($\sim 100-150\; h^{-1}$ Mpc) which is obviously a real feature
of the primordial CDP spectrum. (Contrary to the scale of galaxy
clusters to be a mere consequence of current dynamical time.)
This 'signature of the God' in the primordial spectrum requires
its explanation in physics of the very early Universe.

I cannot help mentioning another connection to the very early
Universe through the primordial perturbation spectrum. This
is a possibility to have high abundance of cosmic
graviotational waves contributing to large-scale CMB
anisotropy. 

There are at least two reasons for such discussion.

The first is theoretical one. Inflation theory is not discriminative to
any of the perturbation modes\cite{15}: both S (CDP) and 
T (CGW) modes can be produced with similar amplitudes
and thus comparable contribution to the CMB anisotropy,
\begin{equation}
\left( \frac{\Delta T}{T}\right)^2_{10^0} = S + T.
\end{equation}

The second reason comes from observations. If the scalar
perturbation spectrum is 'blue' ($n_S > 1$) then the non-zero
T/S is needed to reconcile the COBE $\Delta T/T$ measurement
with the galaxy cluster abundance.

The problem of T/S is
fundamental but can be treated at the moment only theoretically.
A serious discussion on the observational detection of T/S could
be launched after polarization CMB measurements, that would
require the instrumental sensitivity $\sim 1 \mu K$ currently
non-reachable.

\section{Conclusions and Tendencies}

As never before, the cosmologists are very close today to
recover the real model of our Universe and the
post-recombination CDP spectrum directly from observations, both
$\Delta T/T$ and LSS, and make exciting link to the very early
Universe physics. We are going to get data from the advanced
ground and space based CMB explorers as well as huge surveys of
spatial distribution of galaxies, to delimit the cosmological
model with unprecedented precision.

The list of current conclusions may not be full:
\begin{itemize}
\item
extreme open models ($\Omega_M < 0.3$) are rejected by CMB and 
cluster evolution data;
\item
from point of view of distant supernovae Ia the vacuum density of 
our Universe may not be dynamically important ($\Omega_\Lambda < 0.5$);
\item
current data on the Doppler peak indicate small $H_0$ ($\sim 60\; 
km\;s^{-1}$Mpc${}^{-1}$) and large $\Omega_b$ ($\sim 0.1$) and 
$\Omega_M$ ($> 0.5$);
\item
the S-mode fundamental spectrum is consistent with HZ ($n_S\simeq 1-1.2$);
\item
CMB together with LSS data indicate the CDP spectrum break at scale 
$\sim 150$ Mpc, which demands {\it new physical} explanation;
\item
the T/S problem cannot be ignored and needs a careful treatment.
\end{itemize}

\section{Acknowledgements}
The work was partially supported by RFBR (96-02-16689-a). The 
author is grateful to the Organizing Committee for the 
hospitality.


\end{document}